\documentclass[aps,prd,twocolumn,superscriptaddress,floatfix]{revtex4-2}

\usepackage{graphicx}
\usepackage{amsmath,amssymb,amsfonts}
\usepackage{comment}
\usepackage{float}
\usepackage{placeins}
\usepackage[colorlinks,
	linkcolor=blue,
	citecolor=blue,
	urlcolor=blue]{hyperref}
\usepackage{cleveref}
\usepackage{siunitx}
\usepackage[latin1]{inputenc}
\usepackage{bm}
\usepackage{physics}

\newcommand{\shiki}[1]{Eq.~(\ref{#1})}
\newcommand{\zu}[1]{Fig.~{\ref{#1}}}
\newcommand{\hyou}[1]{Table~{\ref{#1}}}
\newcommand{\shou}[1]{Section~{\ref{#1}}}

\newcommand{\gag}{g_{a \gamma \gamma}}
\newcommand{\vvir}{v_{\mathrm{vir}}}

\begin{document}

\begin{flushleft}
{\rm RESCEU-4/23}
\end{flushleft}

\title{First Results of Axion Dark Matter Search with DANCE}

\author{Yuka Oshima}
\email{yuka.oshima@phys.s.u-tokyo.ac.jp}
\affiliation{Department of Physics, University of Tokyo, Bunkyo, Tokyo 113-0033, Japan}
\author{Hiroki Fujimoto}
\affiliation{Department of Physics, University of Tokyo, Bunkyo, Tokyo 113-0033, Japan}
\author{Jun'ya Kume}
\affiliation{Department of Physics, University of Tokyo, Bunkyo, Tokyo 113-0033, Japan}
\affiliation{Research Center for the Early Universe (RESCEU), University of Tokyo, Bunkyo, Tokyo 113-0033, Japan}
\author{Soichiro Morisaki}
\affiliation{Institute for Cosmic Ray Research, University of Tokyo, Kashiwa, Chiba 277-8582, Japan}
\author{Koji Nagano}
\affiliation{Institute of Space and Astronautical Science, Japan Aerospace Exploration Agency, Sagamihara, Kanagawa 252-5210, Japan}
\author{\\Tomohiro Fujita}
\affiliation{Waseda Institute for Advanced Study, Waseda University, Shinjuku, Tokyo 169-8050, Japan}
\affiliation{Research Center for the Early Universe (RESCEU), University of Tokyo, Bunkyo, Tokyo 113-0033, Japan}
\author{Ippei Obata}
\affiliation{Kavli Institute for the Physics and Mathematics of the Universe (WPI), University of Tokyo, Kashiwa, Chiba 277-8583, Japan}
\author{Atsushi Nishizawa}
\affiliation{Research Center for the Early Universe (RESCEU), University of Tokyo, Bunkyo, Tokyo 113-0033, Japan}
\author{Yuta Michimura}
\affiliation{LIGO Laboratory, California Institute of Technology, Pasadena, California 91125, USA}
\affiliation{Research Center for the Early Universe (RESCEU), University of Tokyo, Bunkyo, Tokyo 113-0033, Japan}
\affiliation{PRESTO, Japan Science and Technology Agency (JST), Kawaguchi, Saitama 332-0012, Japan}
\author{Masaki Ando}
\affiliation{Department of Physics, University of Tokyo, Bunkyo, Tokyo 113-0033, Japan}
\affiliation{Research Center for the Early Universe (RESCEU), University of Tokyo, Bunkyo, Tokyo 113-0033, Japan}

\date{\today}

\begin{abstract}
Axions are one of the well-motivated candidates for dark matter,
originally proposed 
to solve the strong \textit{CP} problem in particle physics. 
Dark matter Axion search with riNg Cavity Experiment (DANCE)
is a new experimental project to broadly search for axion dark matter
in the mass range of $10^{-17}~\si{eV} < m_a < 10^{-11}~\si{eV}$.
We aim to detect the rotational oscillation
of linearly polarized light
caused by the axion-photon coupling with a bow-tie cavity.
The first results of the prototype experiment, DANCE Act-1,
are reported from a 24-hour observation.
We found no evidence for axions and set 95\% confidence level
upper limit on the axion-photon coupling 
$\gag \lesssim \SI{8e-4}{GeV^{-1}}$ 
in $10^{-14}~\si{eV} < m_a < 10^{-13}~\si{eV}$.
Although the bound did not exceed 
the current best limits,
this optical cavity experiment is the first demonstration
of polarization-based axion dark matter search 
without any external magnetic field.
\end{abstract}

\maketitle

\section{Introduction}

Axions are hypothetical particles 
generated from a pseudo-scalar field
originally proposed 
to solve the strong \textit{CP} problem 
in quantum chromodynamics (QCD)~\cite{PhysRevLett.38.1440}.
This idea is generally called ``QCD axion''.
Moreover, string theory 
predicts a plenitude of axion-like particles
(ALPs)~\cite{PhysRevD.81.123530}.
QCD axions and ALPs are 
one of the well-motivated candidates for dark matter (DM)
because of their small masses and tiny interactions with matter sectors,
and could behave like a non-relativistic classical wave field 
in the history of the universe~\cite{PRESKILL1983127,
ABBOTT1983133, DINE1983137, Arias_2012}. 
Hereafter in this article,
we collectively call them ``axions''.

The conventional way of searching for axions
is to detect a phenomenon where
axions convert into photons under an external magnetic field
and vice versa,
known as the Primakoff effect~\cite{PhysRevLett.51.1415,
PhysRevLett.52.695, PhysRevD.37.1237}.
Astronomical observations are 
useful to probe the axion-photon conversion 
in the (extra)galactic magnetic fields, 
but no strong evidence has been found~\cite{Payez_2015, Reynolds_2020}.
CERN Axion Solar Telescope (CAST) looked for axions
thermally produced in the Sun with a strong dipole magnet 
and set the current limit 
on the axion-photon coupling~\cite{Anastassopoulos2017}.
Some new projects with toroidal coils
probed a tiny oscillatory magnetic field caused 
by axion DM and achieved the competitive limits 
to CAST~\cite{Gramolin2021, PhysRevLett.127.081801}. 

Recently, new experimental approaches 
to search for axions
were proposed that do not need any strong magnetic field 
but use optical cavities instead~\cite{PhysRevLett.102.202001, 
PhysRevD.98.035021, PhysRevLett.121.161301, 
PhysRevD.100.023548, PhysRevD.101.095034, 
PhysRevLett.123.111301, PhysRevD.104.062008,
Michimura_2020, oshima2021dark, Oshima_2021}.
These methods aim to detect
the phase velocity difference between left- and right-handed 
circular polarizations 
caused by a small coupling 
between axions and photons~\cite{PhysRevD.41.1231, PhysRevLett.81.3067}.
Laser interferometers are good at probing
frequencies of $\lesssim \SI{100}{kHz}$ and 
thus they have a good sensitivity 
in the corresponding axion mass region of 
$m_a \lesssim 10^{-10}~\si{eV}$.
A key point in these techniques
is how to cancel the polarization flipping 
due to the reflection on mirrors 
in order to accentuate the axion effect.
One suggestion is to use
a Mach-Zehnder interferometer 
with two cavities and 
a polarizing beam splitter~\cite{PhysRevLett.102.202001},
and another would be to use a Michelson interferometer 
with quarter-wave plates 
inside two arm cavities~\cite{PhysRevD.98.035021}.
The authors of Ref.~\cite{PhysRevLett.121.161301}
came up with an idea 
to use a bow-tie ring cavity:
Dark matter Axion search with riNg Cavity Experiment
(DANCE)~\cite{Michimura_2020, oshima2021dark, Oshima_2021}.
A bow-tie cavity is used to 
enhance the axion signal
without any optics inside the cavity,
which prevents optical elements from
lowering the finesse of the cavity.
The methods of
injecting linearly polarized beam and 
tuning mirror angles~\cite{PhysRevD.100.023548},
and applying squeezed states of light~\cite{PhysRevD.101.095034} 
were also proposed to improve the sensitivity 
over a wide range of axion masses.

In this work, we demonstrated 
the prototype experiment, DANCE Act-1,
and obtained the upper limit on the axion-photon coupling
from a 24-hour observation.
This optical cavity experiment is the first demonstration
of polarization-based axion DM search 
without any external magnetic field.
This article is organized as follows.
\shou{sec:principle} gives
a brief summary of
the rotational oscillation of optical linear polarization
caused by axion DM 
and an expected sensitivity 
obtained by axion signal amplification 
using a bow-tie cavity.
\shou{sec:exp} reports the experimental setup, 
its performance, and data acquisition.
In \shou{sec:analysis} the data analysis and results 
are described.
\shou{sec:causes} discusses 
the causes of sensitivity degradation. 
Finally, we conclude this work in \shou{sec:conclusion}.

\section{Principle and sensitivity}
\label{sec:principle}

In this section,
we briefly revisit the dynamics of 
the axion field,
calculate its oscillation amplitude,
and derive the sensitivity of our experiment
to the axion-photon coupling.
We set the natural unit $\hbar = c = 1$
unless otherwise noted.

Axions couple to photons 
through the Chern-Simons interaction,
\begin{align}
\mathcal{L} &\supset
\frac{\gag}{4} a(t) F_{\mu \nu} \tilde{F}^{\mu \nu} \nonumber \\
&= \gag \dot{a}(t) \epsilon_{ijk} A_i \partial_j A_k
+ \mathrm{(total\,derivative)},
\end{align}
where dot denotes the time derivative,
$\gag$ is the axion-photon coupling constant,
$a(t)$ is the axion field value,
$A_\mu$ is the vector potential, and
$F_{\mu \nu} \equiv 
\partial_\mu A_\nu - \partial_\nu A_\mu$.
$\tilde{F}^{\mu \nu} \equiv 
\epsilon^{\mu \nu \rho \sigma} F_{\rho \sigma}/2$
is its Hodge dual 
defined with the Levi-Civita antisymmetric tensor
$\epsilon^{\mu \nu \rho \sigma}$.
We impose the temporal gauge $A_0 = 0$
and the Coulomb gauge $\nabla \cdot \bm{A} = 0$.
Then the equation of motion can be written as
\begin{align}
\ddot{A}_i - \nabla ^2 A_i + 
\gag \dot{a}(t) \epsilon_{ijk} \partial_j A_k = 0. 
\label{eq:EoM}
\end{align}
We decompose $A_i$ 
into two circular polarization modes
in the Fourier space
\begin{align}
A_i(t, \bm{x}) = 
\sum_{\alpha = \mathrm{L}, \mathrm{R}}
\int^\infty_{-\infty} \frac{\mathrm{d}^3 k}{(2 \pi)^3}
A_\alpha(t, \bm{k}) e_{\alpha, i}(\hat{\bm{k}})
e^{i \bm{k} \cdot \bm{x}},
\label{eq:Fourier}
\end{align}
where the index $\alpha$ of 
$\mathrm{L}, \mathrm{R}$ represents
left- and right-handed photons,
$\bm{k}$ is the wave number vector,
and $e_{\alpha, i}(\hat{\bm{k}})$
is the circular polarization vector.
By substituting \shiki{eq:Fourier}
into \shiki{eq:EoM},
we obtain
the angular frequencies
\begin{align}
\omega^2_{\mathrm{L} / \mathrm{R}}
= k^2 \qty(1 \mp \frac{\gag \dot{a}(t)}{k})
\end{align}
and the phase velocities 
\begin{align}
c_{\mathrm{L} / \mathrm{R}}
\equiv 
\frac{\omega_{\mathrm{L} / \mathrm{R}}}{k}
\simeq 1 \mp \frac{\gag \dot{a}(t)}{2 k}.
\label{eq:velo}
\end{align}

The axion field is expressed
by a periodic oscillation,
\begin{align}
a(t) = a_0 \cos (m_a t + \delta_\tau(t)),
\label{eq:oscillation}
\end{align}
where the axion mass $m_a$ represents the angular frequency.
The frequency of axion mass can be written as
$f_a = m_a / (2 \pi) 
\simeq \SI{2.4}{Hz} \, (m_a / 10^{-14}~\si{eV})$.
The phase factor $\delta_\tau(t)$ can be regarded as 
a constant value within
the constant timescale of axion DM, $\tau$,
expressed as $\tau = 2 \pi / (m_a v^2_a)$,
where $v_a \sim 10^{-3}$ 
is the DM velocity near the Sun.
Plugging \shiki{eq:oscillation}
into \shiki{eq:velo},
we obtain
\begin{align}
c_{\mathrm{L} / \mathrm{R}} &= 
1 \pm \delta c(t),\\
\delta c(t) &\equiv \frac{\gag \sqrt{2 \rho_a}}{2 k}
\sin (m_a t + \delta_\tau(t)),
\label{eq:velodiff}
\end{align}
where $\rho_a = m_a^2 a_0^2 / 2 = \SI{0.4}{GeV/cm^3}$
is the DM energy density in the Sun.

This phase velocity difference 
causes linearly polarized light
to rotate~\cite{PhysRevD.41.1231, PhysRevLett.81.3067}.
Let $\bm{E}(z=0, t) = E_0 e^{i \omega_0 t} \bm{e}_\mathrm{s}$
defined as the s-polarized injection beam 
propagating along the $z$ axis
with angular frequency $\omega_0$
from $z=0$.
The electric field at $z = l$
can be written as
\begin{align}
&\bm{E}(z=l, t) = E_0 e^{i \omega t} 
(\bm{e}_\mathrm{s}\,\bm{e}_\mathrm{p})
\mqty(1 \\
- \delta \theta (l, t)), \\
&\delta \theta(l, t) \equiv 
k_0 \int^t_{t-l} \delta c(t) \mathrm{d}t \nonumber \\ 
&= \frac{\gag \sqrt{2 \rho_a}}{m_a}
\sin (m_a \qty(t - \frac{l}{2}) + \delta_\tau)
\sin (m_a \frac{l}{2}),
\label{eq:rotation}
\end{align}
when assuming $\delta \theta (l, t) \ll 1$.
Here $k_0 = \omega_0 / c$ is the wave number 
without axions.
The plane of linearly polarized light at $z = 0$ 
rotates by $-\delta \theta(l, t)$ at $z = l$.
Small p-polarized sidebands are generated 
from s-polarized carrier beam,
and vice versa, 
in the presence of axions~\cite{PhysRevD.100.023548}.

The sensitivity of DANCE Act-1 can be calculated 
using a method described in Ref.~\cite{Fujimoto_prep}.
We assume that 
s-polarized light is injected into a bow-tie ring cavity,
with an electric field of
$\bm{E}_\mathrm{in}(t) = E_0 e^{i \omega_0 t} \bm{e}_\mathrm{s}$.
The schematic of an experimental setup for DANCE Act-1
is shown in \zu{fig:setup}
and the symbols of the parameters are summarized
in \hyou{tab:parameter}.
Under the assumption that  
mirrors do not have any optical losses,
the electric field of transmitted light is 
estimated as
\begin{align}
\bm{E}_\mathrm{trans}(t) =& 
\frac{\qty(1 - r_{1\mathrm{s}}^2) e^{-i k_0 l_1}}
{\qty(1 - r_{1\mathrm{s}}^2 r_{2\mathrm{s}}^2)
e^{- i k_0 (2 l_1 + 2 l_2)}} \nonumber \\
&\times E_0 e^{i \omega_0 t} 
(\bm{e}_\mathrm{s}\,\bm{e}_\mathrm{p})
\mqty(1 \\
-\delta \phi (t)), \\
\delta \phi (t) \equiv
& \int_{-\infty}^\infty \frac{\mathrm{d} \omega}{2 \pi}
\Tilde{\delta c} (\omega) e^{i \omega t} H_a(\omega),\label{eq:delta_phi}
\end{align}
where $\delta \phi (t)$ is a polarization rotation angle of 
transmitted light.
$H_a (\omega)$ is a transfer function 
from $\delta c (t)$ to $\delta \phi (t)$:
\begin{align}
&H_a (\omega) \equiv
k_0
\sqrt{\frac{1-|r_{1\mathrm{p}}|^2}{1-|r_{1\mathrm{s}}|^2}}
\frac{1}
{i \omega \qty(1 - r_{1\mathrm{p}}^2 r_{2\mathrm{p}}^2
e^{-i \omega (2l_1 + 2l_2)})} \nonumber \\
&\times \left[ 
\qty(1 - e^{-i \omega l_2}) \right. 
\qty(r_{1\mathrm{s}} r_{2\mathrm{s}} r_{1\mathrm{p}} r_{2\mathrm{p}} 
e^{-i \omega l_1} 
+ r_{1\mathrm{p}}^2 r_{2\mathrm{p}}^2 
e^{-i \omega (2l_1 + l_2)}) \nonumber \\
&\left. -\qty(1 - e^{-i \omega l_1}) 
\qty(r_{1\mathrm{s}} r_{2\mathrm{s}}^2 r_{1\mathrm{p}}
+ r_{1\mathrm{s}} r_{1\mathrm{p}} r_{2\mathrm{p}}^2 
e^{-i \omega (l_1 + l_2)}) \right]
.
\end{align}

The optical path length can be effectively increased 
using an optical cavity
and axion signal is accumulated in the cavity.
We set the reflectivity of s-polarization,
$r_{1\mathrm{s}}$ and $r_{2\mathrm{s}}$,
as real numbers,
whereas that of p-polarization,
$r_{1\mathrm{p}}$ and $r_{2\mathrm{p}}$,
are complex numbers and contain the information about 
the difference of the reflective phase shift
between s- and p-polarizations.
For example, $r_{1\mathrm{p}} = |r_{1\mathrm{p}}|
\exp[-2 \pi i  (2 l_1 + 2 l_2) \delta \nu_1 / c ]$
where $\delta \nu_1$ is 
the reflective phase difference
converted to the free spectral range (FSR)
of a bow-tie cavity.
$c / (2 l_1 + 2 l_2)$ represents the FSR 
and its value in DANCE Act-1 is \SI{302}{MHz}.

\zu{fig:transfun} represents
the response function $|H_a (m_a)|$. 
When $\delta \nu_1 = \delta \nu_2 = 0$,
the sensitivity in the low mass region 
($m_a \lesssim 10^{-12}~\si{eV}$)
will be the highest.
When $\delta \nu_1 \neq 0$ or $\delta \nu_2 \neq 0$,
the sensitivity in the low mass region will be lower.
However, the sensitivity is enhanced at the mass
corresponding to the total reflective phase difference
between s- and p-polarizations
$\delta \nu_\mathrm{total}$
since signal sideband is enhanced in a bow-tie cavity.

\begin{figure*}
\begin{center}
\includegraphics[width=0.95\linewidth]{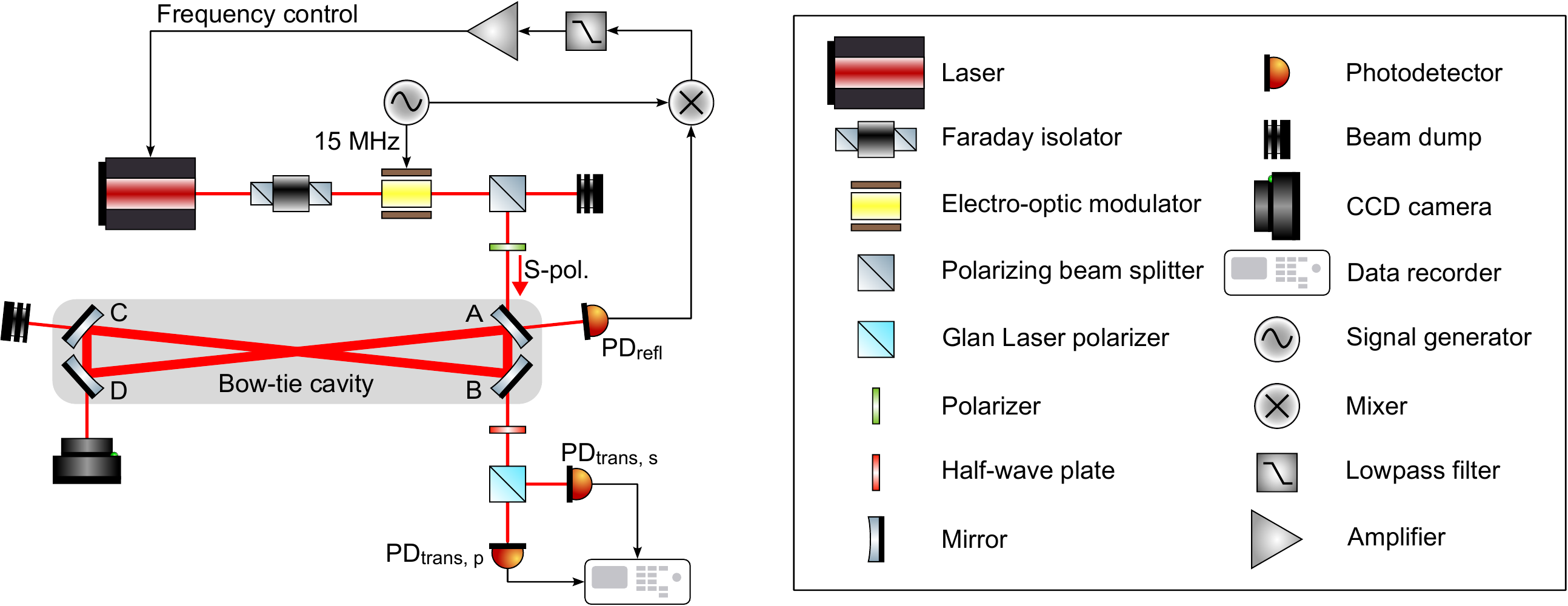}
\caption[]{The schematic of an experimental setup for DANCE Act-1.}
\label{fig:setup}
\end{center}
\end{figure*}

\begin{figure}
\begin{center}
\includegraphics[width=1.0\linewidth]{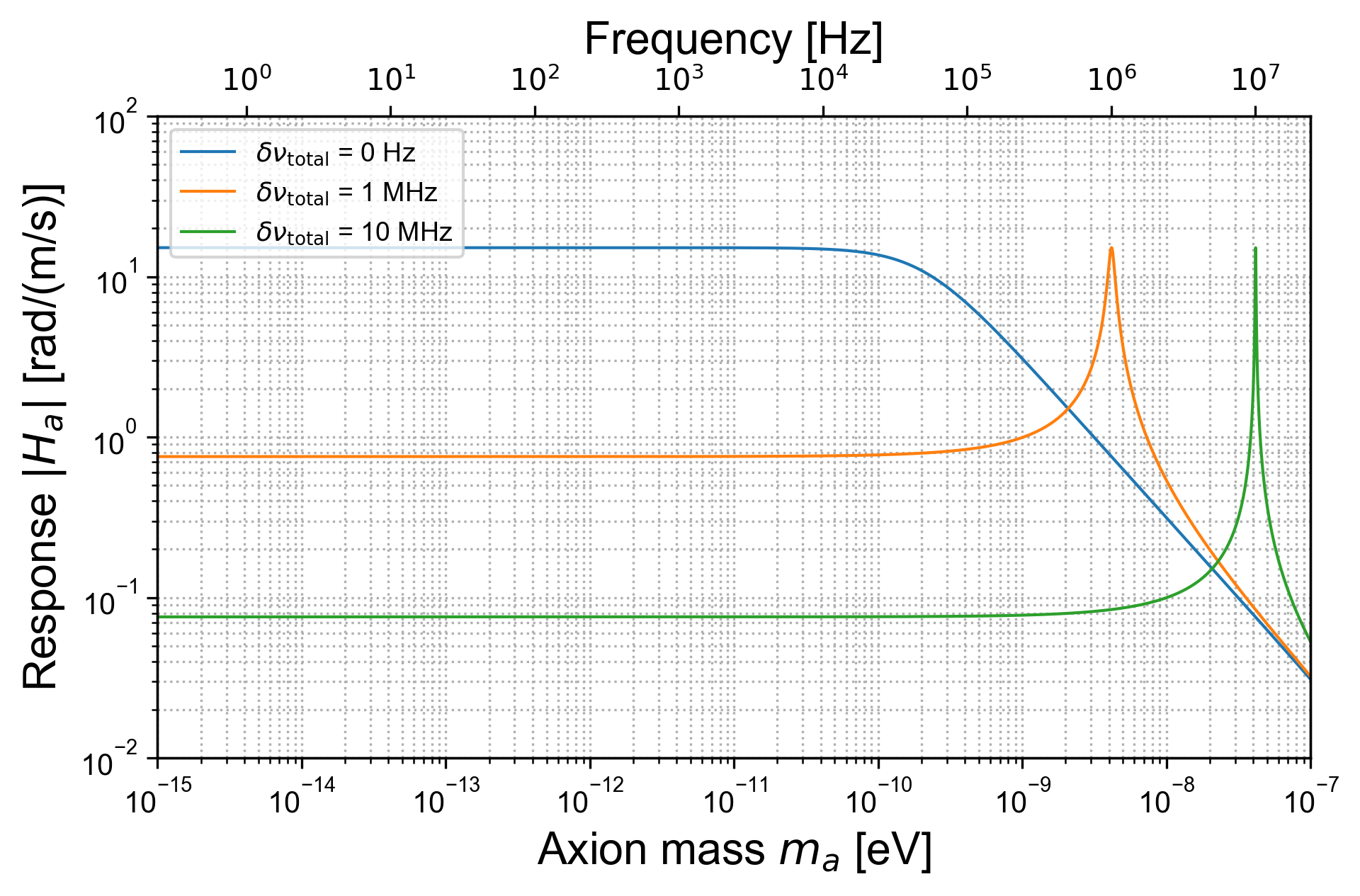}
\caption[]{The response function $|H_a|$.
The blue line shows the function $|H_a|$ for 
$\delta \nu_1 = \delta \nu_2 = \SI{0}{Hz}$
($\delta \nu_\mathrm{total} = \SI{0}{Hz}$),
the orange line for 
$\delta \nu_1 = \delta \nu_2 = \SI{0.25}{MHz}$
($\delta \nu_\mathrm{total} = \SI{1}{MHz}$),
and the green line for 
$\delta \nu_1 = \delta \nu_2 = \SI{2.5}{MHz}$
($\delta \nu_\mathrm{total} = \SI{10}{MHz}$).
The other parameters are the same as 
the final design of DANCE Act-1
as shown in \hyou{tab:parameter}. 
}
\label{fig:transfun}
\end{center}
\end{figure}

The sensitivity of DANCE depends on 
the method of detection.
The detailed setup of this work is described 
in \shou{sec:exp}.
The fundamental noise source of DANCE
would be quantum shot noise and
the potential sensitivity limited by shot noise
is roughly estimated as

\begin{align}
&\gag \geq
\left\{ 
\begin{array}{l}
\SI{227}{GeV^{-1}}
\frac{\qty|1 - r_{1\mathrm{s}}^2 r_{2\mathrm{s}}^2|}
{\qty|1 - r_{1\mathrm{s}}^2|}
\frac{\si{eV^{-1}}}{|H^\prime_a(m_a)|/k_0} \\
\times \sqrt{
\frac{\si{GeV/cm^3}}{\rho_a}
\frac{\si{W}}{P_\mathrm{in}}
\frac{\si{\um}}{\lambda_0}
\frac{\si{s}}{T_\mathrm{obs}}
}
\,\,\,\,\,
(T_\mathrm{obs} \lesssim \tau),
\\
\SI{227}{GeV^{-1}}
\frac{\qty|1 - r_{1\mathrm{s}}^2 r_{2\mathrm{s}}^2|}
{\qty|1 - r_{1\mathrm{s}}^2|}
\frac{\si{eV^{-1}}}{|H^\prime_a(m_a)|/k_0} \\
\times \sqrt{
\frac{\si{GeV/cm^3}}{\rho_a}
\frac{\si{W}}{P_\mathrm{in}}
\frac{\si{\um}}{\lambda_0}
\frac{\si{s}}{(T_\mathrm{obs} \tau)^{1/2}}
} 
\,\,\,\,\,
(T_\mathrm{obs} \gtrsim \tau),
\end{array}
\right. \\
&|H^\prime_a (m_a)| \equiv
\frac{1}{2} \left\{ (\operatorname{Re}
\qty[H_a (m_a) + H_a (-m_a)])^2
\right. \nonumber \\
& \hspace{18mm}
\left. + (\operatorname{Im} 
\qty[H_a (m_a) - H_a
(-m_a)])^2 \right\}^{\frac{1}{2}},
\end{align}
where $\lambda_0$ is a laser wavelength and 
$|H^\prime_a (m_a)|$ is a signal amplification factor 
by a bow-tie cavity.
We assume that we can observe axions when
ratio between axion signal and shot noise $\geq 1$.
The sensitivity improves 
as the measurement time increases,
with the factor of $T_\mathrm{obs}^{1/2}$
as long as the axion oscillation is 
coherent for $T_\mathrm{obs} \lesssim \tau$, 
where $\tau$ is the coherent timescale of axion DM.
When the measurement time becomes longer than 
this coherence time $T_\mathrm{obs} \gtrsim \tau$,
the proportionality of the sensitivity
with the measurement time changes to $(T_\mathrm{obs}\tau)^{1/4}$.
This different proportionality is owing to the stochasticity of the amplitude of axion field~\cite{nakatsuka2022stochastic}, which will be discussed in \shou{sec:analysis}.

Even using conservative parameters listed in \hyou{tab:parameter}, 
DANCE Act-1 can exceed the CAST limit
(see the green curve in \zu{fig:upperlimits}).
If we use more optimistic parameters,
with a
round-trip length of \SI{10}{m},
finesse of $10^6$,
and input laser power of \SI{100}{W},
we can reach $\gag < \SI{3e-16}{GeV^{-1}}$ 
for $m_a < 10^{-16}~\si{eV}$~\cite{PhysRevLett.121.161301}
and improve the sensitivity broadly
by several orders of magnitude
compared to the best upper limits at present.

\begin{table*}
\begin{center}
\caption{Summary of the parameters for DANCE Act-1.
The difference of the reflective phase shift
between s- and p-polarizations
is written in the conversion to the FSR of the cavity
($\delta \nu = \mathrm{phase\,shift\,[rad]} / (2 \pi)
\times c / (2 l_1 + 2 l_2)$).
In the column for \textit{this experiment},
the values with stars * are specifications,
and the others are measured values.
The parameters of the final design are listed
to show the ideal sensitivity 
(plotted as the green curve in \zu{fig:upperlimits}).
}
\begin{ruledtabular}
\begin{tabular}{lccc}
Parameter & Symbol & Final design & This experiment
\\
\hline
Injected laser power &
$P_\mathrm{in}$ & \SI{1}{W} & \SI{242(12)}{mW} \\
Transmitted laser power &
$P_\mathrm{trans}$ & \SI{1}{W} & \SI{153(8)}{mW} \\
Distance between A and D, B and C &
$l_1$ & \SI{45}{cm} & \SI{45}{cm} * \\
Distance between A and B, C and D &
$l_2$ & \SI{4.7}{cm} & \SI{4.7}{cm} * \\
Power reflectivity of s-pol. (A and B) &
$|r_{1 \mathrm s}|^2$ & 99.9\% & 99.90(2)\% * \\
Power reflectivity of s-pol. (C and D) &
$|r_{2 \mathrm s}|^2$ & 100\% & $>$99.99\% * \\
Power reflectivity of p-pol. (A and B) &
$|r_{1 \mathrm p}|^2$ & 99.9\% & 98.42(2)\% \\
Power reflectivity of p-pol. (C and D) &
$|r_{2 \mathrm p}|^2$ & 100\% & 99.95(1)\% \\
Finesse of s-pol. (carrier) &
$\mathcal{F}_\mathrm{s}$ & \num{3e3} & \num{2.85(5)e3} \\
Finesse of p-pol. (signal sidebands) &
$\mathcal{F}_\mathrm{p}$ & \num{3e3} & 195(3) \\
Total difference of the reflective phase shift between s- and p-pol. &
$\delta \nu_\mathrm{total}$ & \SI{0}{Hz} & \SI{2.52(2)}{MHz} \\
Difference of the reflective phase shift between s- and p-pol.
(A and B) &
$\delta \nu_1$ & \SI{0}{Hz} & \SI{-0.55(97)}{MHz} \\
Difference of the reflective phase shift between s- and p-pol.
(C and D) &
$\delta \nu_2$ & \SI{0}{Hz} & \SI{2.08(99)}{MHz}
\end{tabular}
\end{ruledtabular}
\label{tab:parameter}
\end{center}
\end{table*}

\section{Experiment}
\label{sec:exp}

In this section, the experimental setup, its performance
and data acquisition are reported.

\subsection{Setup and performance}

\zu{fig:setup} shows the experimental setup of DANCE Act-1.
We used a Nd:YAG laser, Mephisto 500 NE, 
with a wavelength of 1064 nm.
The s-polarized beam
was injected into a bow-tie cavity.
We put a polarizing beam splitter
as well as a polarizer in front of the cavity
to have linearly polarized light injected into the cavity.
Our bow-tie cavity was constructed from four mirrors A-D
rigidly fixed on a spacer made of aluminum.

We aim to probe the axion signal 
by taking the interference
between a carrier beam (s-polarization in this work)
and signal sidebands (p-polarization)
in the direction of 
amplitude quadrature~\cite{PhysRevD.100.023548}.
Polarization of transmitted light was rotated 
with a half-wave plate
to introduce some p-polarized reference signal
which has the same frequency as a carrier beam,
and then split
into s- and p-polarizations with a Glan Laser polarizer.
The amplitudes of s- and p-polarizations were monitored
with photodetectors PD$_\mathrm{trans, s}$ and PD$_\mathrm{trans, p}$
and saved with a data recorder for two weeks.

The laser frequency was locked to the resonance of TEM00 mode
by obtaining the error signal for the laser frequency control
with the Pound-Drever-Hall method~\cite{Drever1983}.
Spatial mode is confirmed to be TEM00 using a CCD camera.
To improve the lock duration time,
the double-loop feedback control system and 
the automated cavity locking system were developed
~\cite{fujimoto2021dark}.
Feedback signal above $\sim \SI{30}{Hz}$ 
was sent into the laser fast port (piezo actuator),
and feedback signal under $\sim \SI{30}{Hz}$ 
was sent into the laser slow port (temperature actuator).
To implement this system, we used SEAGULL mini 
as a digital signal processor, 
and also as a lowpass filter for the low frequency control loop.
A digital signal processor monitored the output of PD$_\mathrm{trans, s}$
and identified whether the cavity was locked or unlocked.
When the cavity was unlocked, 
signal into the laser slow port was swept 
until the cavity is locked again.

$l_1$ was designed to be around 10 times longer than $l_2$
to enhance the rotational oscillation of s-polarization
by preventing the linear polarization from inverting 
when reflecting on mirrors.
We specified only $|r_{1 \mathrm s}|^2$ and $|r_{2 \mathrm s}|^2$
when we ordered custom-made mirrors A-D
because it is difficult to control 
the reflective phase shift
and to satisfy our requirements. 
All the four mirrors were concave mirrors 
with a radius of curvature of \SI{1}{m}.
Beam diameter on the mirrors was $\sim \SI{800}{\micro m}$.
Incident angles at all the four mirrors were \SI{42}{deg}.

$\mathcal{F}_\mathrm{s}$, $\mathcal{F}_\mathrm{p}$, and
$\delta \nu_\mathrm{total}$
were measured 
by sweeping cavity resonances.
$\mathcal{F}_\mathrm{s}$ was consistent 
with the specified reflectivity
$|r_{1 \mathrm s}|^2$ and $|r_{2 \mathrm s}|^2$
and we could achieve a high finesse.
$\delta \nu_\mathrm{total}$ was non-zero
because each polarization obtains a different phase shift
from mirror-coating layers
when reflecting at oblique incident angles.
Note that $\delta \nu_\mathrm{total}$ drifted 
from \SI{2.52(2)}{MHz} to \SI{0.50(2)}{MHz}
in the two-week observation.
We obtained $\delta \nu_1$ and $\delta \nu_2$ 
separately for data analysis.
We used mirrors with different coating layers 
to build the cavity
and measured $\delta \nu_\mathrm{total}$ 
with various mirror combinations.
Assuming that mirrors with the same coating layers 
have the same phase shift,
we determined $\delta \nu_1$ and $\delta \nu_2$.

\subsection{Data acquisition}

The time series data of s- and p-polarizations,
$P_\mathrm{s}(t)$ and $P_\mathrm{p}(t)$,
was observed 
with a sampling rate of \SI{1}{kHz}
for 1,004,400 seconds
in May 18-30, 2021.
We analyzed two sets of continuous 
86,400-second (24-hour) data on May 18 and 19
because the first two days 
were the stretch of time with the most stable lock.
One set was used to set the upper limit
and the other was used to veto candidate peaks.

We calibrated the output of photodetectors 
$P_\mathrm{s}(t)$ and $P_\mathrm{p}(t)$
to the rotation angle of linear polarization $\phi (t)$ by 
\begin{align}
\phi(t) = \sqrt{\frac{P_\mathrm{p}(t)}
{P_\mathrm{s}(t) + P_\mathrm{p}(t)}} - 2\theta_\mathrm{HWP},
\label{eq:rotangle}
\end{align}
where $\theta_\mathrm{HWP}$ is the rotation angle 
of the half-wave plate
with respect to the s-polarized light
at the detection port.
We do not need to measure $\theta_\mathrm{HWP}$
because it is a constant parameter
and we focus on oscillational amplitudes.
The one-sided amplitude spectral density (ASD) of the observed rotation angle of linear polarization is plotted in \zu{fig:ASD}.
We reached \SI{3.4e-6}{rad/\sqrt{Hz}} at \SI{5}{Hz}.

\begin{figure}
\begin{center}
\includegraphics[width=1.0\linewidth]{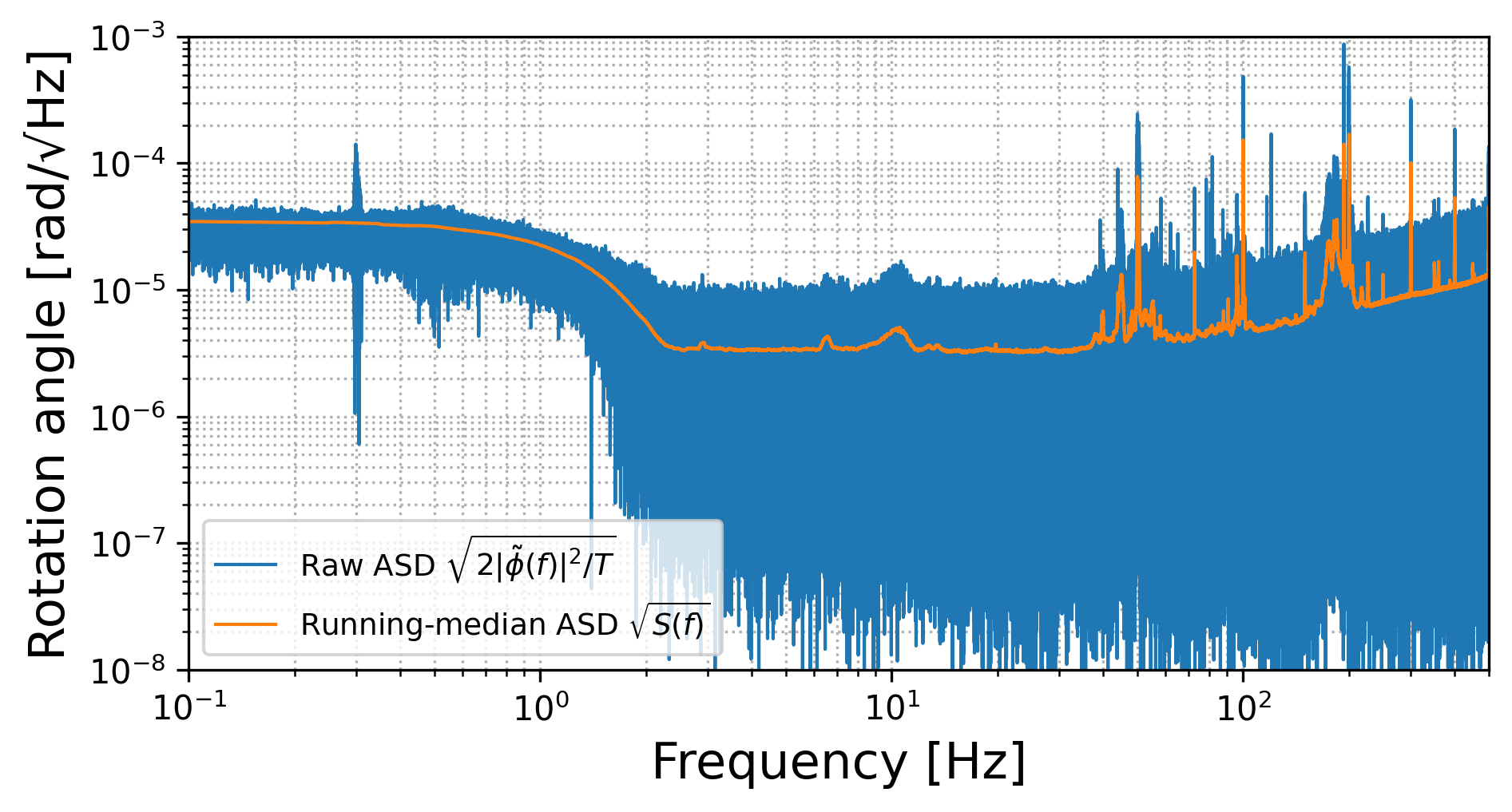}
\caption[]{The one-sided ASD 
of the rotation angle of linear polarization.}
\label{fig:ASD}
\end{center}
\end{figure}

\section{Data analysis}
\label{sec:analysis}

In this section, we describe our analysis of the data
to place upper limit 
on the axion-photon coupling.

\subsection{Detection statistics}

The signal is expected to have a bandwidth of $\sim f_a \vvir^2$, 
where $\vvir$ is the virial velocity 
of our Galaxy~\cite{PhysRevA.97.042506}. 
Thus, for axion mass value $m_a$, we define the following signal-to-noise ratio (SNR) as a detection statistic:
\begin{align}
\rho \equiv \sum_{f_a \leq f_n \leq f_a(1 +\kappa^2 \vvir^2)}\frac{4|\tilde{\phi}(f_n)|^2}{TS(f_n)}
,\label{det_stat}
\end{align}
where $T$ is the duration of the data segment, 
$f_n$ is the discretized frequency bin, 
$\kappa$ is a constant of order unity,
$\tilde{\phi}(f)$ represents the Fourier-transformation of $\phi (t)$, 
and $S(f)$ is the one-sided noise power spectral density. 
Note that in the absence of the signal $\delta\phi(t)$,
this corresponds to the orange curve in~\zu{fig:ASD}.
Raw ASD is also plotted in~\zu{fig:ASD}.
We adopt 
$\vvir=\SI{220}{km/s}$~\cite{BERTONE2005279, PhysRevD.99.023012}, 
and $\kappa=3.17$ to guarantee 
that the fractional loss of signal is 
less than $99\%$ assuming the standard halo model of 
DM velocity distribution.
Here $S(f_n)$ is evaluated by the running median 
from $\sim 8,600$ neighboring frequency bins 
in order to smear out the effect of DM signal 
localized in the narrow band. 

The detection threshold of $\rho$ is determined 
under the assumption that the instrumental noise 
is a stationary Gaussian process. 
In the absence of a signal, 
$\rho$ follows a $\chi^2$ distribution 
with $2N_\mathrm{bin}$ degrees of freedom, 
where $N_\mathrm{bin}$ denotes the number of frequency bins 
involved in the sum of \shiki{det_stat}.
We chose the threshold 
to be the $95\%$ percentile of that distribution. 
For each case where the measured value of $\rho$ exceeds this threshold, 
we performed the veto analysis as explained below.

The upper bound on the signal amplitude 
is calculated in the frequentist method introduced 
by Ref.~\cite{nakatsuka2022stochastic}. 
Because the axion field is superposition of particle waves with random phase, its amplitude  randomly fluctuates. 
The analysis method proposed by the previous work takes into account this random axion amplitude.
Let $\delta\phi^{\rm (upp)}(m_a)$ denote the upper bound on the root-mean-square (RMS) of $\delta\phi(t)$ in Eq.~\eqref{eq:delta_phi} for axion mass value $m_a$.
At the confidence level $\beta$, 
it is calculated by the following equation,
\begin{align}
1-\beta = 
\int_0^{\rho_{\rm mea}(m_a)}{\rm d} \rho~\mathcal{L}
(\rho|\delta\phi^{\rm (upp)}(m_a)),\label{eq:frequentist}
\end{align}
where $\rho_{\rm mea}(m_a)$ is the measured value of $\rho$,
and $\mathcal{L}(\rho|\delta \phi)$ is the likelihood 
of observing detection statistics $\rho$ 
conditioned on signal with RMS $\delta \phi$. 
Note again that the effect of randomness 
in the axion DM amplitude mentioned above is included in the likelihood function derived in Ref.~\cite{nakatsuka2022stochastic}.
The interested readers can find the concrete expressions of this likelihood in Appendix.~\ref{sec:intermediate_results}.
We chose $\beta = 0.95$ for the numerical calculation of the upper bound.
This upper bound on the rotation angle 
can be converted into that on the axion-photon coupling 
through the following relation:
\begin{align}
&\gag(m_a) = \frac{2 k_0 \delta \phi^{\rm (upp)}(m_a)}
{\sqrt{2 \rho_a}|H^\prime_a (m_a)|} \nonumber \\
&= \SI{5.10e11}{GeV^{-1}} 
\frac{\si{eV^{-1}}}{|H^\prime_a(m_a)|/k_0}
\sqrt{
\frac{\si{GeV/cm^3}}{\rho_a}
}
\delta\phi^{\rm (upp)}(m_a)
.
\label{eq:gag}
\end{align}

\subsection{Results}

After passing the first 24-hour data set through our pipeline for calculating $\rho_{\rm mea}(m_a)$,
556 points exceeded 
the detection threshold of $\rho$ out of a total of 1,776,390 points in \num{0.1}-\SI{490}{Hz} as shown in \zu{fig:SNR}.
We conducted the following two veto procedures:
the persistence veto and the linewidth veto.

An axion signal should have the same frequency
in two segments of data
with the accuracy of 
$\Delta \omega / \omega \sim 10^{-6}$~\cite{PhysRevA.97.042506}.
We rejected the points that did not match
the second set of data
with 6 significant digits accuracy.
This persistence veto reduced 
the number of candidate points to 257.

Since the expected linewidth of the galactic DM
is $\Delta \omega / \omega \sim 10^{-6}$~\cite{PhysRevA.97.042506},
we eliminated the points that formed 
a peak wider than $10^{-5}$.
The candidate points were decreased to 7 by this linewidth veto.

The frequencies of remaining peaks 
are summarized in \hyou{tab:SNR}.
All the peaks were approximately multiples of \SI{40}{Hz}.
As you can see in \zu{fig:PDHpeak},
peaks in the error signal of the laser frequency control
had the same frequency 
as the peaks that were not rejected in the veto process.
As the axion signal should not be present in the error signal, 
this suggests that the cause of remaining candidate peaks 
are from mechanical resonances of the cavity.
We therefore rejected all the remaining candidate peaks.

\begin{figure}
\begin{center}
\includegraphics[width=1.0\linewidth]{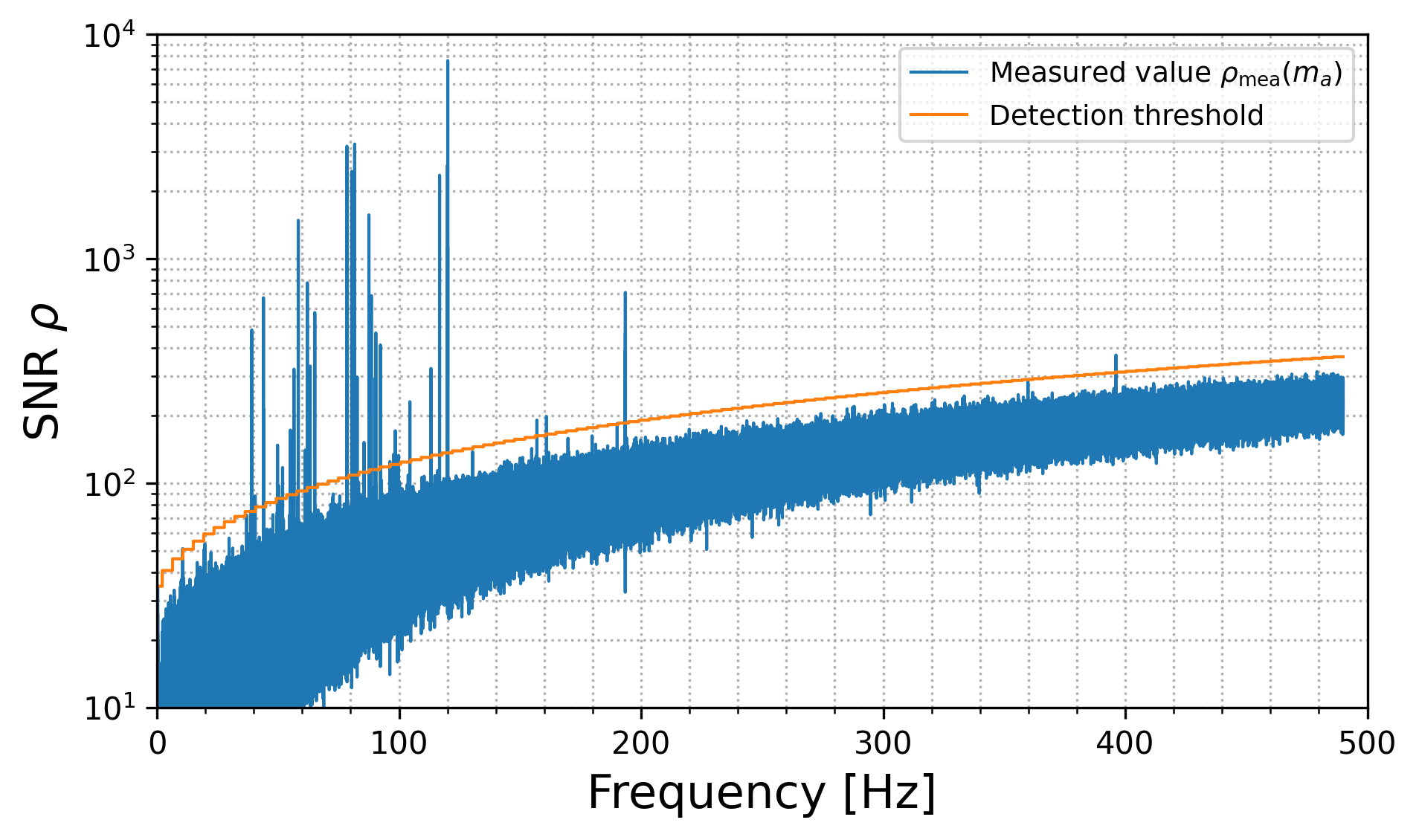}
\caption[]{SNR $\rho$ of axion DM signal for each frequency $f_a$.
The blue line shows the measured value $\rho_{\rm mea}(m_a)$ and
the orange line represents the detection threshold.}
\label{fig:SNR}
\end{center}
\end{figure}

\begin{table}
\begin{center}
\caption{Summary of remaining peaks after the veto procedures.}
\begin{ruledtabular}
\begin{tabular}{ccc}
Frequency & 
\begin{tabular}{c}
SNR $\rho$\\(Measured value)
\end{tabular}
&
\begin{tabular}{c}
SNR $\rho$\\(Detection threshold)
\end{tabular}
\\
\hline
\SI{81.6712}{Hz} & 3243 & 109 \\
\SI{119.983}{Hz} & 2073 & 137 \\
\SI{120.001}{Hz} & 2616 & 137 \\
\SI{120.113}{Hz} & 1125 & 137 \\
\SI{120.117}{Hz} & 159 & 137 \\
\SI{120.118}{Hz} & 7637 & 137 \\
\SI{396.142}{Hz} & 373 & 313 \\
\end{tabular}
\end{ruledtabular}
\label{tab:SNR}
\end{center}
\end{table}

\begin{figure}
\begin{center}
\includegraphics[width=1.0\linewidth]{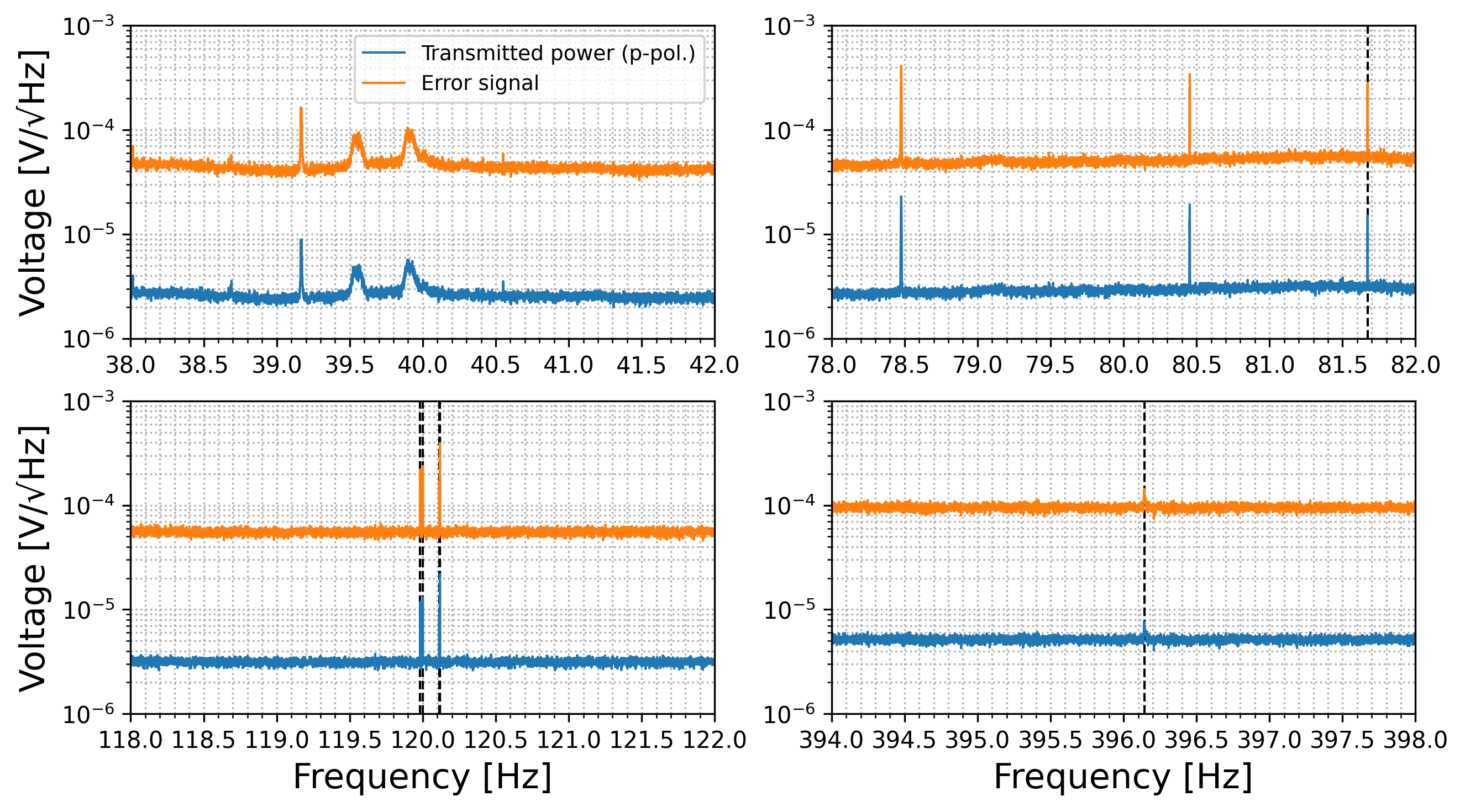}
\caption[]{The one-sided amplitude spectral density 
of the transmitted p-polarization signal (blue line)
and the error signal (orange line)
around \SI{40}{Hz} (upper left panel),
\SI{80}{Hz} (upper right panel), 
\SI{120}{Hz} (lower left panel),
and \SI{400}{Hz} (lower right panel). 
Black dashed lines are the frequencies
corresponding to the 7 remaining peaks
summarized in \hyou{tab:SNR}.}
\label{fig:PDHpeak}
\end{center}
\end{figure}

We obtained the spectrum of the upper limit 
to the rotation angle of linearly polarized light
$\delta\phi^{\rm (upp)}(m_a)$
from the analysis pipeline,
and calibrated it to the bound
on the axion-photon coupling $\gag(m_a)$
from \shiki{eq:gag}.
Note that 
the upper bound on the rotation angle
of linear polarization $\delta\phi^{\rm (upp)}(m_a)$
and the response function $|H^{\prime}_a|$ with the parameters of this work are plotted in 
Appendix~\ref{sec:intermediate_results}.

The initial value of 
$\delta \nu_\mathrm{total} = \SI{2.52(2)}{MHz}$ and
the combination of $\delta \nu_1$ and $\delta \nu_2$
were found to be a major source of systematic effect,
which will give 11\% of difference in the upper limit 
at $m_a = 10^{-13}\,{\rm eV}$. 
The values of these parameters were chosen 
to set the most conservative upper limit.
The results are shown in \zu {fig:upperlimits}.
The upper limit was limited by classical noises
and worse than 
the current shot noise by 5 orders of magnitude.
Since $\delta \nu_1$ and $\delta \nu_2$ were non-zero,
the current shot noise sensitivity 
was worse than the design sensitivity
in the low mass range and 
has the dip at $m_a \simeq 10^{-8}~\si{eV}$
which corresponds to the frequency of $\delta \nu_\mathrm{total}$.

\begin{figure}
\begin{center}
\includegraphics[width=1.0\linewidth]{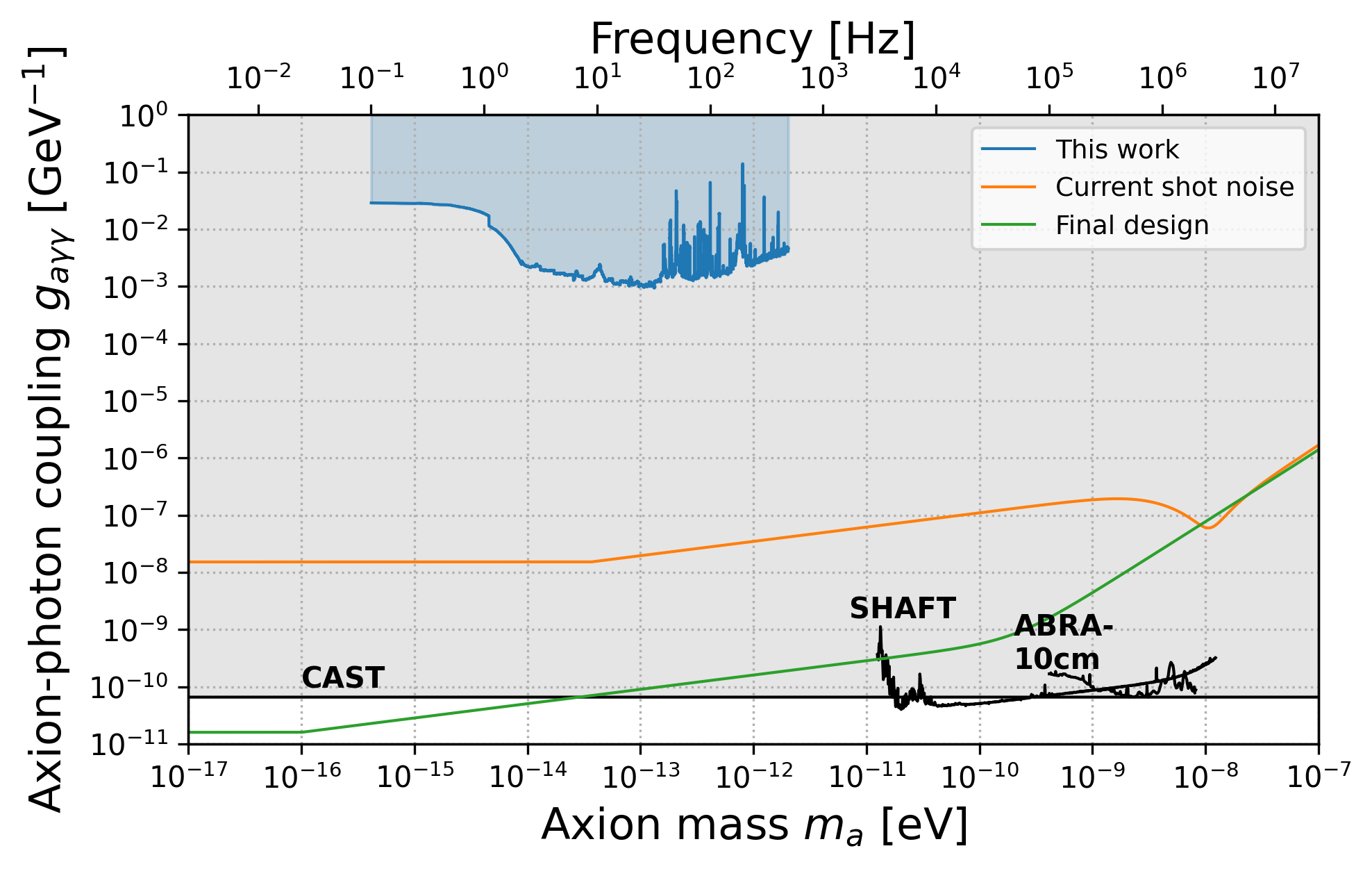}
\caption[]{
The blue line shows the upper limit
on the axion-photon coupling constant 
obtained by this work.
We observed for $T_\mathrm{obs} = $~86,400$~\si{seconds}$.
The orange line is the expected sensitivity
limited by shot noise
with the same parameters as this work 
assuming $T_\mathrm{obs} = $~86,400$~\si{seconds}$.
The green line
represents the designed shot noise limited sensitivity
of DANCE Act-1 
assuming a observation time of $T_\mathrm{obs} = \SI{1}{year}$.
The parameters which are used for this estimation
are summarized in \hyou{tab:parameter}.
The black lines with the grey-shaded region 
are current bounds obtained from
CAST~\cite{Anastassopoulos2017}, SHAFT~\cite{Gramolin2021},
and ABRACADABRA-10cm~\cite{PhysRevLett.127.081801} experiments.
}
\label{fig:upperlimits}
\end{center}
\end{figure}

\section{Causes of sensitivity degradation}
\label{sec:causes}

We discuss the two causes of sensitivity degradation
in this experiment here.
One is classical noise sources
and the other is a non-zero phase difference 
between the two polarizations.

The rotation angle of linear polarization
in \num{0.1}-\SI{1}{Hz}
correlated significantly
with the injected laser power,
and the rotation angle of linear polarization
in \SI{30}{Hz}-\SI{5}{kHz} correlated
with the error signal for the frequency control.
Thus, laser intensity noise, laser frequency noise, and 
mechanical vibration
are some of the candidates for noise sources
limiting our sensitivity.
Furthermore, in principle the phase noises 
such as laser frequency noise and mechanical vibration
are not supposed to contribute to noise for DANCE, 
which observes the signal in amplitude quadrature,
but it could have coupled in this demonstration.
The reduction of these noise sources is underway in our upgraded setup 
to be reported in future work.

The sensitivity is also reduced
because of the reflective phase difference
between two linear polarizations.
If we can realize $\delta \nu_1 = \delta \nu_2 = 0$,
the sensitivity will improve by 3 orders of magnitude.
We aim to deal with this issue 
by constructing an auxiliary cavity 
to achieve simultaneous resonance 
between both polarizations~\cite{PhysRevD.101.095034, Fujimoto_2021}.

\section{Conclusion}
\label{sec:conclusion}

The broadband axion DM search
with a bow-tie cavity, DANCE Act-1 was demonstrated.
We searched for the rotation and oscillation
of linearly polarized light
caused by the axion-photon coupling 
for 86,400 seconds
and obtained the first results by DANCE.
We found no evidence for axions and set 95\% confidence level
upper limit on the axion-photon coupling 
$\gag \lesssim \SI{8e-4}{GeV^{-1}}$ 
in the axion mass range of 
$10^{-14}~\si{eV} < m_a < 10^{-13}~\si{eV}$.

The candidates for noise sources limiting our sensitivity 
are laser intensity noise, frequency noise, and mechanical vibration.
The sensitivity will be improved
by introducing laser intensity control and 
a vibration isolation system
as well as upgrading the frequency control system.
The difference of reflective phase shift between
s- and p-polarizations is also the cause
for the sensitivity degradation.
We are installing an auxiliary cavity 
to realize simultaneous resonance 
between the two polarizations~\cite{Fujimoto_2021}.

Although the upper limit did not exceed 
the current best limits,
this optical cavity experiment is the first demonstration
of polarization-based axion dark matter search 
without any external magnetic field.
By sufficiently upgrading the setup
using the techniques mentioned above,
we are expecting to improve the sensitivity 
by several orders of magnitude.

\section{Acknowledgement}

We would like to thank 
Shigemi Otsuka and Togo Shimozawa
for manufacturing the mechanical parts,
Kentaro Komori and Satoru Takano 
for fruitful discussions, 
and Ching Pin Ooi for editing this paper.
This work is supported by JSPS KAKENHI Grant Nos. 
JP18K13537, JP20H05850, JP20H05854, and JP20H05859, 
by the Sumitomo Foundation,
and by JST PRESTO Grant No. JPMJPR200B.
Y.~O. is supported by Grant-in-Aid for JSPS Fellows No. JP22J21087
and by JSR Fellowship, the University of Tokyo.
H.~F. is supported by Grant-in-Aid for JSPS Fellows No. JP22J21530 
and by Forefront Physics and Mathematics Program 
to Drive Transformation (FoPM),
a World-leading Innovative Graduate Study (WINGS) Program,
the University of Tokyo.
J.~K. is supported by 
Grant-in-Aid for JSPS Fellows No. JP20J21866 
and by research program of the Leading Graduate Course 
for Frontiers of Mathematical Sciences and Physics (FMSP).
A.~N. is supported by JSPS KAKENHI Grant Nos. JP19H01894 
and JP20H04726 and by Research Grants from Inamori Foundation.
I.~O. is supported by JSPS KAKENHI Grant No. JP19K14702.

\appendix
\section{Likelihood function of SNR and the upper limits}
\label{sec:intermediate_results}

As shown in Ref.~\cite{nakatsuka2022stochastic}, the Fourier mode of signal with mass $m_a$ can be expressed in terms of the stochastic variables as
\begin{align}
    \delta\phi(f_n ; m_a)
    &\simeq
    \gag(m_a)\frac{\sqrt{2 \rho_a}}{2 k_0}|H^\prime_a (m_a)|T\\
    &~~\times\sqrt{ \Delta_s(f_n ; m_a)}
    ~
    \left[
    \frac{r_n }{{\sqrt{2}}}
    e^{i\theta_n}
    \right],\notag\\
    &\equiv \delta\phi(m_a)\frac{T}{2}
    \sqrt{ \Delta_s(f_n ; m_a)}
    \left[
    \frac{r_n }{{\sqrt{2}}}
    e^{i\theta_n}
    \right]
    ,
    \label{eq_signal_axions}
\end{align}
where $\theta_n$ and $r_n$ respectively obey a uniform distribution over $\left[0, 2\pi\right]$ and the standard Rayleigh distribution.
$\Delta_s(f_n; m_a)$ is the analytic function that represents the deterministic part of the spectral shape determined by the velocity distribution of DM (see Ref.~\cite{nakatsuka2022stochastic} for details).
By performing the marginalization over $r_n$, the likelihood that takes into account the random amplitude of DM signal can be obtained as
\begin{align}
    \mathcal{L}(\rho_n|\lambda_n)
    & \equiv 
    \int  {\rm  d}r_n~
    P_R (r_n)
     {\mathcal L}\left (\rho_n \big | \lambda_n r_n \right)
\nonumber\\&=
     \frac{1}{{2}(1+\lambda_n^2)} \exp\left( 
     \frac{-\rho_n}{{2}(1+\lambda_n^2)}
     \right)
     ,
     \label{eq_barL_single}
\end{align}
where $\rho_n = 4|\tilde{\phi}(f_n)|^2/TS(f_n)$ is the SNR at frequency bin $f_n$. As can be seen, the likelihood is characterized by the parameter
\begin{align}
    \lambda_n \equiv \delta\phi(m_a)\sqrt{\Delta_s(f_n ; m_a)}\sqrt{\frac{T}{2S(f_n)}},
\end{align}
that depends on the characteristic amplitude of signal $\delta\phi(m_a)$.
Then from this expression, the likelihood for the (total) SNR $\rho = \Sigma_n\rho_n$ defined in Eq.~\eqref{det_stat} can be derived as
 \begin{align}
    \mathcal{L}(\rho|\{\lambda_n\})
	&=
	\int
	\left(
	\prod_{l }^{N_\mathrm{bin}}
	{\rm d} \rho_n~
	\mathcal{L} (\rho_n|\lambda_n)
	\right)
	\delta\left(\rho - \sum_n^{N_\mathrm{bin}} \rho_n  \right)
 \nonumber\\& =
    \sum_n^{N_\mathrm{bin}}  
     \frac{
    w_n
     }{{2}(1+\lambda_n^2)}
      \exp\left( 
     - \frac{\rho}{{2}(1+\lambda_n^2)}
     \right)
	,\label{eq:full_likelihood}
	\\
    w_n
	&\equiv
	\prod_{n'(\neq n)}^{N_\mathrm{bin}}
	\frac{1+\lambda_n^2}{\lambda_n^2-\lambda_{n'}^2},
\end{align}
where $\lambda_n\neq \lambda_{n'}$ is assumed for all frequency bins $n\neq n'$. We should note that this assumption would be violated and there arises a numerical instability, specifically for higher DM masses which involves more frequency bins in Eq.~\eqref{det_stat}.
In this case, however, we can use the Gaussian approximation of Eq.~\eqref{eq:full_likelihood}:
\begin{align}
	{\mathcal L}(\rho|\{\lambda_n\})
	&\to 
	\frac{1}{\sqrt{2\pi \sigma^2_\rho } }
	\exp\left(
	-
	\frac{(\rho - \mu_\rho)^2}{2\sigma^2_\rho}
	\right)
	\quad\mathrm{for}\quad
    T\gg\tau,
    \label{eq_approx_likelihood}
    \\
    \mu_\rho
    &=
    2\sum_n  (1+\lambda_n^2),
    \\
     \sigma^2_\rho
    &=
    {4}\sum_n (1+\lambda_n^2)^2
    .
\end{align}
In our analysis, we apply this Gaussian approximation for $N_\mathrm{bin} > 8$.

From these expressions of likelihood function, we could numerically set the 95\% confidence limit on $\left\{\lambda_n\right\}$, or equivalently on $\delta\phi(m_a)$ according to Eq.~\eqref{eq:frequentist}.
The upper bound $\delta\phi^{\rm (upp)}(m_a)$ derived in our pipeline is shown in \zu{fig:upperlimits_rad}.
As mentioned in the main text, $\delta\phi^{\rm (upp)}(m_a)$ is converted to the upper limit on $\gag$ with \shiki{eq:gag}. This was achieved by using the response function $|H^{\prime}_a|$ presented in \zu{fig:Ha_exp}.

\begin{figure}[H]
\begin{center}
\includegraphics[width=1.0\linewidth]{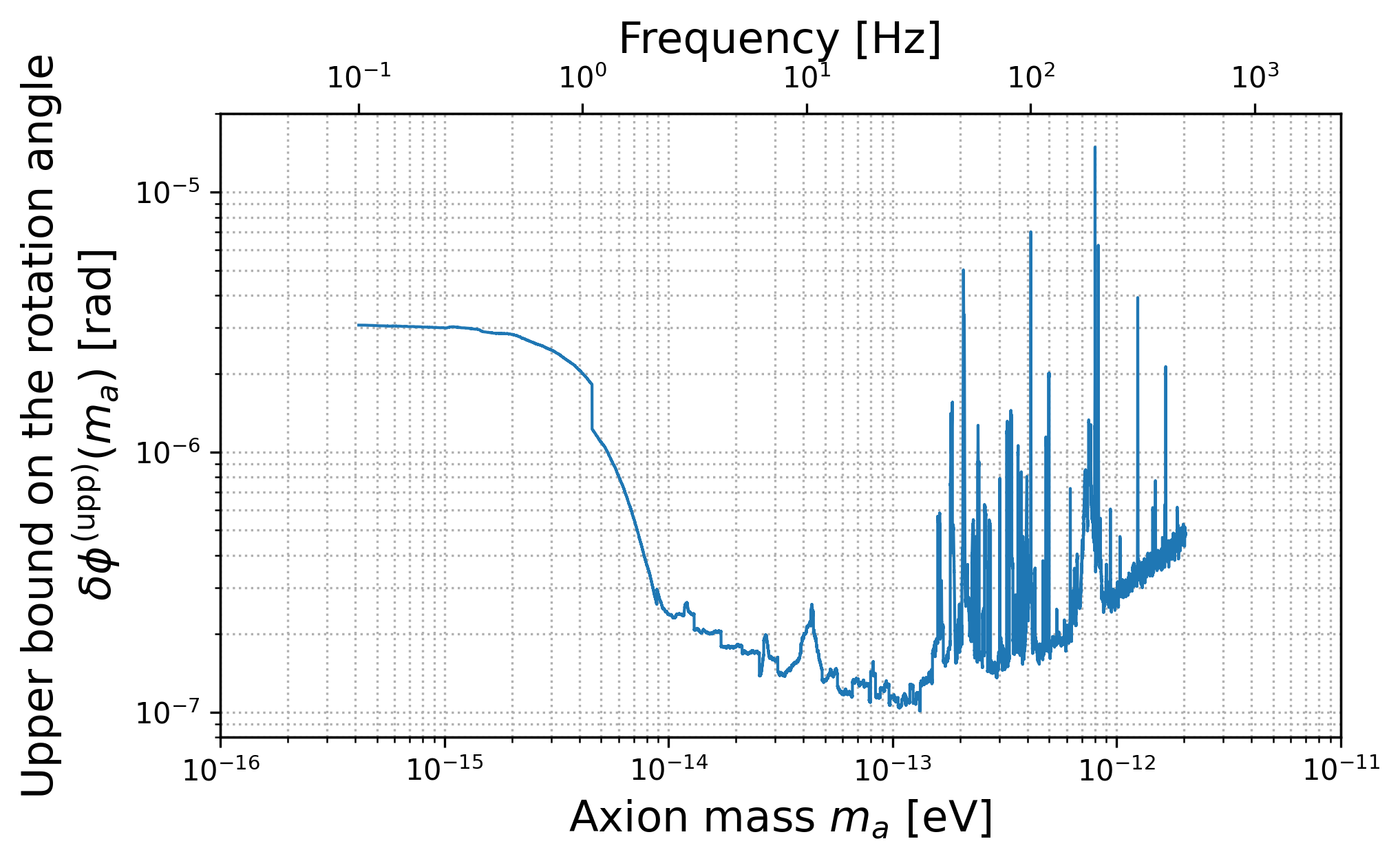}
\caption[]{The upper bound on the rotation angle
of linear polarization $\delta\phi^{\rm (upp)}(m_a)$.}
\label{fig:upperlimits_rad}
\end{center}
\end{figure}

\begin{figure}[H]
\begin{center}
\includegraphics[width=1.0\linewidth]{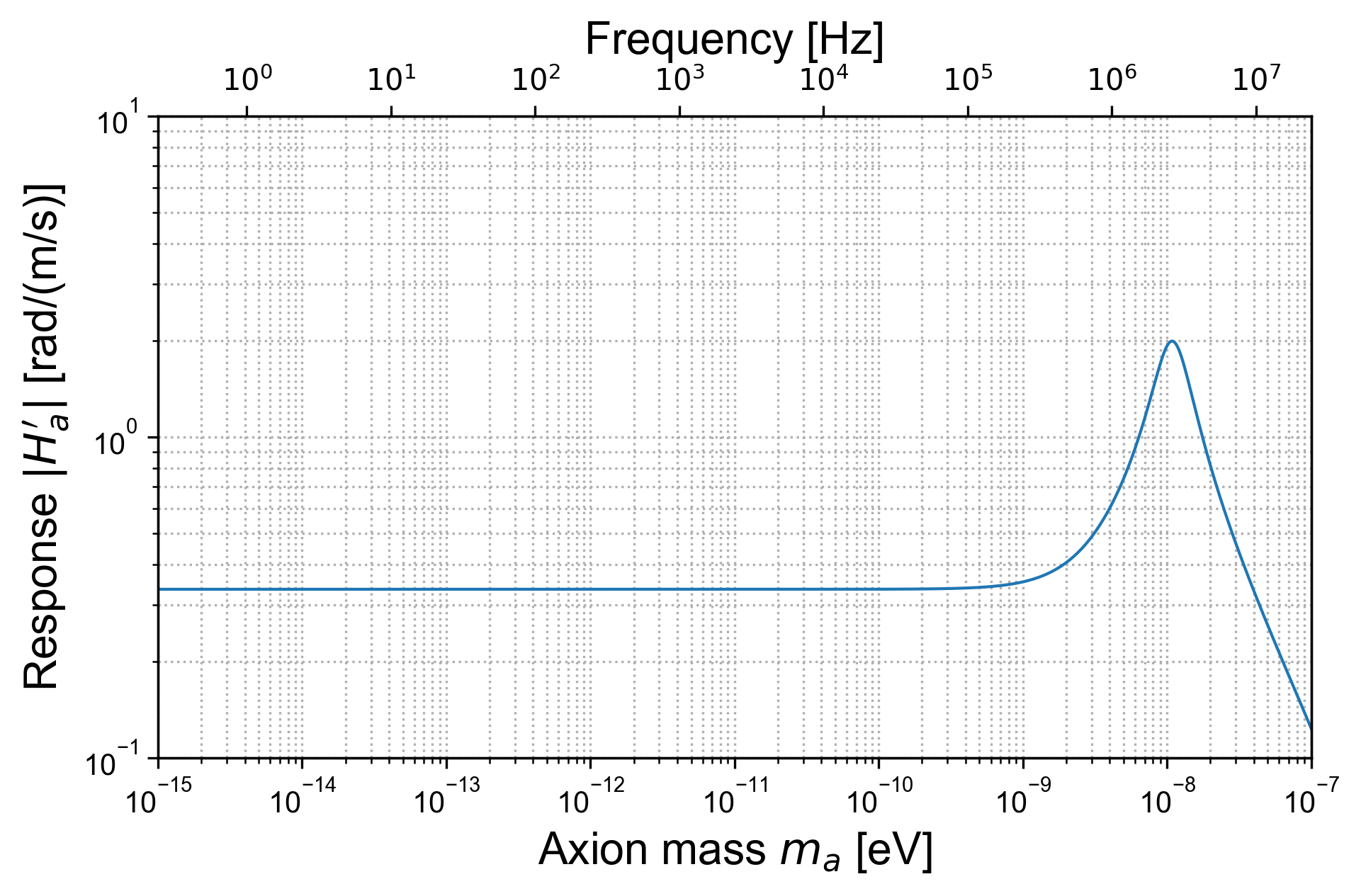}
\caption[]{The response function
with the parameters of this experiment $|H^{\prime}_a|$.
The parameters
were chosen from \hyou{tab:parameter} to set the most conservative upper
limit.}
\label{fig:Ha_exp}
\end{center}
\end{figure}

\bibliography{DANCEFirstResults}
\bibliographystyle{apsrev.bst}
\end{document}